\documentclass[11pt,a4paper]{article}
\usepackage[T1]{fontenc}
\usepackage[utf8]{inputenc}
\usepackage{amsmath,amssymb,amsthm}
\usepackage[margin=2.5cm]{geometry}
\usepackage{hyperref}

\newtheorem{theorem}{Theorem}

\title{Riemann spaces associated with the Navier--Stokes equations:\\
a survey of the geometric approach}
\author{Valerii S. Dryuma\\[2pt]
\small Vladimir Andrunachievici Institute of Mathematics and Computer Science,\\
\small Moldova State University, Academiei str.~5, MD-2028 Chi\c{s}in\u{a}u, Republic of Moldova\\
\small \texttt{valdryum@gmail.com} \quad ORCID: 0009-0002-8591-6650}
\date{}

\begin{document}
\maketitle

\begin{abstract}
We survey a geometric approach to the Navier--Stokes equations developed by the
author over the past fifteen years. The central object is a Riemannian space of
fourteen dimensions whose metric is Ricci-flat precisely on the solutions of the
Navier--Stokes system, so that the behaviour of a viscous incompressible fluid
becomes a property of curvature. The space decomposes into a flat six-dimensional
part and two dual quadruples of coordinates, Eulerian and Lagrangian, which
exhibits the two classical descriptions of a fluid as two halves of a single
geometry. We also recall the associated six-dimensional metric, whose
integrability condition is the incompressibility of the fluid; the link with the
projective geometry of second-order ordinary differential equations in the sense
of E.~Cartan; and the use of Cartan's invariants to construct a three-dimensional
analogue of the Taylor--Green vortex. An application of the same construction to
the equations of rotation of a rigid body is indicated.
\end{abstract}



\section{Introduction}

The properties of the flow of a viscous incompressible fluid are described by the
Navier--Stokes system
\begin{equation}
\begin{aligned}
&U_t + U U_x + V U_y + W U_z - \mu\,(U_{xx}+U_{yy}+U_{zz}) + P_x = 0,\\
&V_t + U V_x + V V_y + W V_z - \mu\,(V_{xx}+V_{yy}+V_{zz}) + P_y = 0,\\
&W_t + U W_x + V W_y + W W_z - \mu\,(W_{xx}+W_{yy}+W_{zz}) + P_z = 0,\\
&U_x + V_y + W_z = 0,
\end{aligned}
\label{eq:ns}
\end{equation}
where $U,V,W$ are the components of the velocity of the flow, $P$ is the pressure
and $\mu$ is the coefficient of viscosity, all functions of $(x,y,z,t)$. The
construction of exact solutions of this system of four nonlinear partial
differential equations remains an important open problem.

The approach surveyed here rests on the following idea. One associates with
\eqref{eq:ns} a Riemannian space whose metric is Ricci-flat exactly on the
solutions of the system. Properties of the flow then become properties of the
curvature of that space, and the tools of Riemannian and projective differential
geometry --- geodesics, Killing vectors, the invariants of E.~Cartan --- become
available for the study of the flow.

The basis of the geometric description is the pair of relations
\[
x = f_1(a,b,c,t),\quad y = f_2(a,b,c,t),\quad z = f_3(a,b,c,t),\quad
p = h(a,b,c,t),
\]
where $(x,y,z,p)$ are local coordinates for the pressure and the position, and
$(a,b,c)$ are the dual coordinates of the individual points of the liquid at each
instant $t$. Eliminating $(a,b,c)$ by appropriate differentiations gives the
Navier--Stokes system in the Euler variables; eliminating $(x,y,z)$ instead gives
the system governing the motion of the individual points of the liquid.

\section{The 14-dimensional Ricci-flat metric}

\begin{theorem}
The metric of the $14$-dimensional space in local coordinates
$(x,y,z,t,\eta,\rho,m,u,v,w,p,\xi,\chi,n)$
\begin{equation}
\begin{aligned}
ds^2 = {}& 2\,dx^2 + 2\,dx\,dy + 2\,dx\,du + 2\,dy^2 + 2\,dy\,dz + 2\,dy\,dv\\
&+ 2\,dz^2 + 2\,dz\,dw + 2\,dt\,dp + 2\,d\eta\,d\xi + 2\,d\rho\,d\chi
+ 2\,dm\,dn\\
&+ A\,dt^2 + B\,d\eta^2 + C\,d\rho^2 + E\,dm^2,
\end{aligned}
\label{eq:14d}
\end{equation}
where
\begin{align}
A &= 2 - U(x,y,z,t)\,u - V(x,y,z,t)\,v - W(x,y,z,t)\,w,\\
B &= \bigl(-UW + \mu\,U_z\bigr)w + \bigl(-UV + \mu\,U_y\bigr)v
 + \bigl(\mu\,U_x - U^2 - P\bigr)u - U p,\\
C &= \bigl(-VW + \mu\,V_z\bigr)w + \bigl(\mu\,V_y - V^2 - P\bigr)v
 + \bigl(-UV + \mu\,V_x\bigr)u - V p,\\
E &= \bigl(-\mu\,U_x - \mu\,V_y - W^2 - P\bigr)w
 + \bigl(-VW + \mu\,W_y\bigr)v + \bigl(-UW + \mu\,W_x\bigr)u - W p,
\end{align}
is Ricci-flat on the solutions of the Navier--Stokes system \eqref{eq:ns}.
\end{theorem}

The space so obtained consists of a flat six-dimensional part in the coordinates
$(x,y,z,u,v,w)$ together with two dual four-dimensional subspaces, one in the
Euler coordinates and one in the Lagrangian coordinates. The two classical
descriptions of a fluid thus appear as two halves of one geometry, and the
structure of the partition depends essentially on the properties of the flat
six-dimensional part, itself divided into two three-dimensional subspaces.

The metric \eqref{eq:14d} belongs to the class of partially projective Riemannian
spaces whose scalar invariants vanish; spaces of this type also occur in the
theory of gravitational waves. For their study one may use either the invariants
of E.~Cartan or the classical Beltrami--Laplace invariants.

\section{The system in Lagrangian variables}

In the Lagrangian variables $(a,b,c,t)$ the full Navier--Stokes system takes the
considerably more complicated form
\begin{align}
&\frac{\partial^2 A}{\partial t^2} + [B,C,P]
- \mu\bigl([B,C,[B,C,A_t]] + [C,A,[C,A,A_t]] + [A,B,[A,B,A_t]]\bigr) = 0,\\
&\frac{\partial^2 B}{\partial t^2} + [C,A,P]
- \mu\bigl([B,C,[B,C,B_t]] + [C,A,[C,A,B_t]] + [A,B,[A,B,B_t]]\bigr) = 0,\\
&\frac{\partial^2 C}{\partial t^2} + [A,B,P]
- \mu\bigl([B,C,[B,C,C_t]] + [C,A,[C,A,C_t]] + [A,B,[A,B,C_t]]\bigr) = 0,\\
&[A,B,C] - 1 = 0 .
\end{align}

\section{The 6-dimensional metric and incompressibility}

\begin{theorem}
The $6$-dimensional Riemann space in the coordinates $(x,y,z,a,b,c,t)$ equipped
with the metric
\begin{equation}
\begin{aligned}
ds^2 = {}& A(a,b,c,t)\,dx^2 + 2B(a,b,c,t)\,dx\,dy + 2E(a,b,c,t)\,dx\,dz + dx\,da\\
&+ C(a,b,c,t)\,dy^2 + 2H(a,b,c,t)\,dy\,dz + dy\,db\\
&+ F(a,b,c,t)\,dz^2 + dz\,dc
\end{aligned}
\end{equation}
can be used for the integration of the equation
\begin{equation}
\begin{aligned}
&B_b\,C_c\,A_a - B_b\,C_a\,A_c - C_c\,B_a\,A_b - C_b\,B_c\,A_a\\
&\quad + C_b\,B_a\,A_c + B_c\,C_a\,A_b - 1 = 0,
\end{aligned}
\end{equation}
which is the condition of incompressibility of the liquid.
\end{theorem}

In a particular case the metric simplifies and the solutions of the
Navier--Stokes system are expressed through a function $P(x)$ satisfying the
Monge--Amp\`ere equation.

\section{Projective duality and second-order ODEs}

The geometric method for studying the Navier--Stokes system is connected with the
theory of second-order ordinary differential equations of the form
\begin{equation}
y'' + a_1(x,y)\,(y')^3 + 3a_2(x,y)\,(y')^2 + 3a_3(x,y)\,y' + a_4(x,y) = 0
\label{eq:ode}
\end{equation}
in the sense of the normal projective connection of E.~Cartan. The general
integral of \eqref{eq:ode} has the form $H(x,y,a,b)=0$ in four variables, whence
follows a differential equation in the variables $a,b$,
\[
\frac{d^2 b}{da^2} = g\!\left(a,b,\frac{db}{da}\right).
\]

\begin{theorem}
The two equations
$y'' = -a_1 y'^3 - 3a_2 y'^2 - 3a_3 y' - a_4$ and $b'' = g(a,b,b')$, where
$b' = c$, form a dual pair with a common integral $H(x,y,a,b)=0$, determined with
the help of the solutions of the system
\begin{equation}
\begin{aligned}
h &= g_{ca} + g\,g_{cc} - \tfrac12\,(g_c)^2 + c\,g_{cb} - 2 g_b,\\
0 &= h_{ca} + g\,h_{cc} - g_c\,h_c + c\,h_{cb} - 3 h_b .
\end{aligned}
\end{equation}
\end{theorem}

\section{The 4-dimensional metric and its geodesics}

\begin{theorem}
The geodesic equations of the $4$-dimensional space with the metric
\begin{equation}
\begin{aligned}
ds^2 = {}& \bigl(2 z a_3(x,y) - 2 t a_4(x,y)\bigr)dx^2
 + 2\bigl(2 z a_2(x,y) - 2 t a_3(x,y)\bigr)dx\,dy\\
&+ 2\,dx\,dz + \bigl(2 z a_1(x,y) - 2 t a_2(x,y)\bigr)dy^2 + 2\,dy\,dt
\end{aligned}
\end{equation}
in the coordinates $(x,y,z,t)$ decompose into two parts. The first consists of the
two nonlinear equations
\begin{align}
&\ddot y + a_4(x,y)\,\dot x^2 + 2a_3(x,y)\,\dot x\,\dot y + a_2(x,y)\,\dot y^2 = 0,\\
&\ddot x - a_3(x,y)\,\dot x^2 - 2a_2(x,y)\,\dot x\,\dot y - a_1(x,y)\,\dot y^2 = 0,
\end{align}
the dot denoting differentiation with respect to the parameter $s$. The second is
the linear system
\begin{equation}
\frac{d^2\vec\Psi}{ds^2} + A(x,y)\,\frac{d\vec\Psi}{ds} + B(x,y)\,\vec\Psi = 0,
\qquad \vec\Psi = (\Psi_1 = z(s),\ \Psi_2 = t(s)),
\end{equation}
with $2\times2$ matrix-functions $A(x,y)$, $B(x,y)$.
\end{theorem}

The Ricci tensor of this metric is
\begin{equation}
\begin{aligned}
R_{11} &= 2\,\partial_y a_4 - 2\,\partial_x a_3 - 4 a_3^2 + 4 a_4 a_2,\\
R_{12} &= -2\,\partial_x a_2 + 2\,\partial_y a_3 - 2 a_2 a_3 + 2 a_4 a_1,\\
R_{22} &= -2\,\partial_x a_1 + 2\,\partial_y a_2 + 4 a_1 a_3 - 4 a_2^2 .
\end{aligned}
\end{equation}

\section{Killing vectors and the invariants of E.~Cartan}

In the construction of solutions of the Navier--Stokes equations, both smooth and
singular, one uses the Killing equations
\begin{equation}
K_{i,j} + K_{j,i} - 2\Gamma^k_{ij}K_k = 0,
\qquad\text{equivalently}\qquad
K^k g_{ij,k} + g_{ik}K^k_{,j} + g_{jk}K^k_{,i} = 0,
\end{equation}
together with the Lie derivatives of the corresponding vector fields $u^k$,
\begin{equation}
u^i_{j,k} + u^n\Gamma^i_{jk,n} + u^n_{,j}\Gamma^i_{nk}
+ u^n_{,k}\Gamma^i_{jn} - u^n_{,n}\Gamma^i_{jk} = 0,
\end{equation}
where $\Gamma^i_{jk}$ are the connection coefficients of the metric
$ds^2 = g_{ij}\,dx^i dx^j$.

As an example of the use of the non-vanishing invariants of E.~Cartan we recall
the construction of a three-dimensional analogue of the classical two-dimensional
Taylor--Green vortex
\begin{equation}
\begin{aligned}
&U = \cos x\,\sin y\;e^{-2\mu t}, \qquad V = -\sin x\,\cos y\;e^{-2\mu t},\\
&W = 0, \qquad P = -\tfrac14\bigl(\cos 2x + \cos 2y\bigr)e^{-4\mu t}.
\end{aligned}
\end{equation}
For a vector with components $A^k = [a,b,c,e,0,0,0,f,l,m,n,0,0,0]$ depending on
$(x,y,z,t)$, the invariant of E.~Cartan built from the Riemann curvature tensor
$R^i{}_{jkl}$ of the metric takes the form
\begin{equation}
T = V W_y + U W_x + (V_y)^2 + U V_y + U_y V_x + (U_x)^2 .
\end{equation}
Solving the Navier--Stokes equations with account of the invariant $T$ leads to
new examples of three-dimensional flows of Taylor--Green type.

\section{Geodesics of the 14-dimensional space}

The geodesic lines of the metric \eqref{eq:14d} are the solutions of a system of
four second-order nonlinear ordinary differential equations in the coordinates
$(x,y,z,t)$ together with four second-order linear ordinary differential
equations for the dual coordinates $(u,v,w,p)$, with coefficients depending on
the components of the curvature tensor of the space. The six additive coordinates
are flat and form a configuration of six straight lines as geodesics.

An important role in the theory of integration of the Navier--Stokes equations
belongs to the conditions of their compatibility. These admit a geometric
description based on the properties of a six-dimensional space equipped with a
Riemannian metric subject to special conditions on its components. The
coordinates $(x,y,z,t)$ and $(u,v,w,p)$ are dual to each other, and their
properties are determined by the geometry of spaces of normal projective
connectivity in the sense of E.~Cartan for the corresponding pair of coordinates.

From various conditions on the curvature tensors of the metric and on the
equations of the geodesic lines one obtains the equation of a hypersurface
\begin{equation}
\Psi(x,y,z,t,u,v,w,p) = 0,
\end{equation}
which is of importance for the understanding of the topological properties of the
Navier--Stokes equations.

\section{Application to the rotation of a rigid body}

With the help of the Ricci-flat metric of the fourteen-dimensional space, the
solutions of the Navier--Stokes system have been applied to the theory of the
rotation of a rigid body, the properties of the Kovalevskaya top being considered
in more detail.

\section{Conclusion}

The geometric approach surveyed here proposes a way of constructing exact
solutions of the Navier--Stokes equations and of studying their properties, by
transferring the question from the analysis of a nonlinear system of partial
differential equations to the geometry of an associated Riemannian space. The
same machinery applies to the Euler equations, to the Kadomtsev--Petviashvili
equation and to the equations of rotation of a rigid body.

\section*{Acknowledgements}

The work was supported in part by the Institutional Research Programme
011303 ``SATGED'' of Moldova State University and by the National Agency for
Research and Development of Moldova (grant 2170086.31SD).

\end{document}